\documentclass[runningheads]{llncs}
\usepackage{graphicx}
\begin{document}
\title{An Integrated Web Platform for the Mizar Mathematical Library}
%
%
\author{Hideharu Furushima\inst{1} \and
Daichi Yamamichi\inst{1} \and
Seigo Shigenaka\inst{1} \and
Kazuhisa Nakasho\inst{1,3}\orcidID{0000-0003-1110-4342} \and
Katsumi Wasaki\inst{2}}
\authorrunning{H. Furushima et al.}
%
\institute{Yamaguchi University, Yamaguchi, Japan
\email{\{b089vgv,a102vgu,a085vgu,nakasho\}@yamaguchi-u.ac.jp} \and
Shinshu University, Nagano, Japan
\email{\{wasaki\}@cs.shinshu-u.ac.jp} \and
corresponding author
}
\maketitle              
\begin{abstract}
This paper reports on the development of a Web platform to host the Mizar Mathematical Library (MML). In recent years, the size of formalized mathematical libraries has been drastically increasing, and this has led to a growing demand for tools that support efficient and comprehensive browsing, searching, and annotation of these libraries. This platform implements a Wiki function to add comments to the HTMLized MML, three types of search function (article, symbol, and theorem), and a function to show the dependency graph of the MML. This platform is designed with consistency, scalability, and interoperability as top priorities for long-term use.

\end{abstract}
\section{Introduction}
Many systems have been proposed to improve the browsability and searchability of formalized mathematical libraries. The MathWiki Project\footnote{{https://www.nwo.nl/en/projects/612066825}} aims to improve the readability of formalized mathematical libraries and make them accessible to wider communities. A Wiki for Mizar~\cite{urban2010wiki}, Large Formal Wikis for Coq/CoRN~\cite{alama2011large}, and Agora System for Flyspeck Project~\cite{tankink2013formal} were proposed in the MathWiki project. However, since the MathWiki Project was terminated in 2014, its contents cannot follow the libraries' updates. ProofWiki~\cite{vyskocil2018disambiguating} and the Lean Mathematical Library~\cite{van2020maintaining} accumulate mathematical libraries and convert them into highly readable HTML documents. However, these two systems do not collaborate with advanced search engines or graphical tools to visualize library dependencies.

We developed the emwiki system, a Wiki service that hosts the Mizar Mathematics Library (MML)~\cite{bancerek2018role} while addressing issues with existing systems. The emwiki system is a Web platform based on the Django framework\footnote{https://www.djangoproject.com/}, featuring extensibility and interoperability. Vue.js\footnote{https://vuejs.org/} is used as the front-end framework, making it a partial single-page application. This eliminates extra rendering to reduce user stress. Vuetify\footnote{https://vuetifyjs.com/} is used as the user interface library, which allows for intuitive operation. Currently, this service is available at {https://em1.cs.shinshu-u.ac.jp/emwiki/release/}.

\section{Wiki Function}
The Wiki feature is implemented to embed additional comments to read and understand articles in the MML.
Although users cannot edit mathematical statements written in the formal language itself, they can add comments to theorems and definitions.
We reused the HTMLized MML~\cite{urban2013atp} to improve the convenience of the MML.
The HTMLized MML is a document in which the MML is converted to HTML format, and the reference relationships are expressed as hyperlinks.
We adopted the LaTeX syntax for mathematical expressions in the comments, which is familiar to mathematicians.
MathJax~\cite{cervone2012mathjax} is used as a mathematical expression rendering engine.
A real-time preview feature is also implemented to check rendering results while editing.

The version tracking feature manages the history of comments.
This function is achieved by using Git, a distributed version control system, as its backend.
Moreover, it is necessary to maintain the linkage between theorems/definitions and comments during library updates.
Our system exploits the Git merge function for this purpose.
As theorems/definitions do not have persistent identifiers in the Mizar language, the most effective way to identify theorems/definitions before and after a library update is to track text differences.
Our system embeds comments directly into the MML.
Since the comments are written just before the theorems/definitions, administrators can maintain the consistency of the linkage between the theorems/definitions and comments using the three-way merge function during library updates.

The user management function allows administrators to track and block malicious users.
The Wiki function stores a history of editors and their revisions, and when a comment is accidentally rewritten, it is possible to contact the user who wrote it and roll back it.
Also, if a malicious comment is found, it will be deleted, and the user who wrote it will be blocked.
The user management function is available in any component on our platform.

\begin{figure}[hbt!]
        \includegraphics[width=\textwidth]{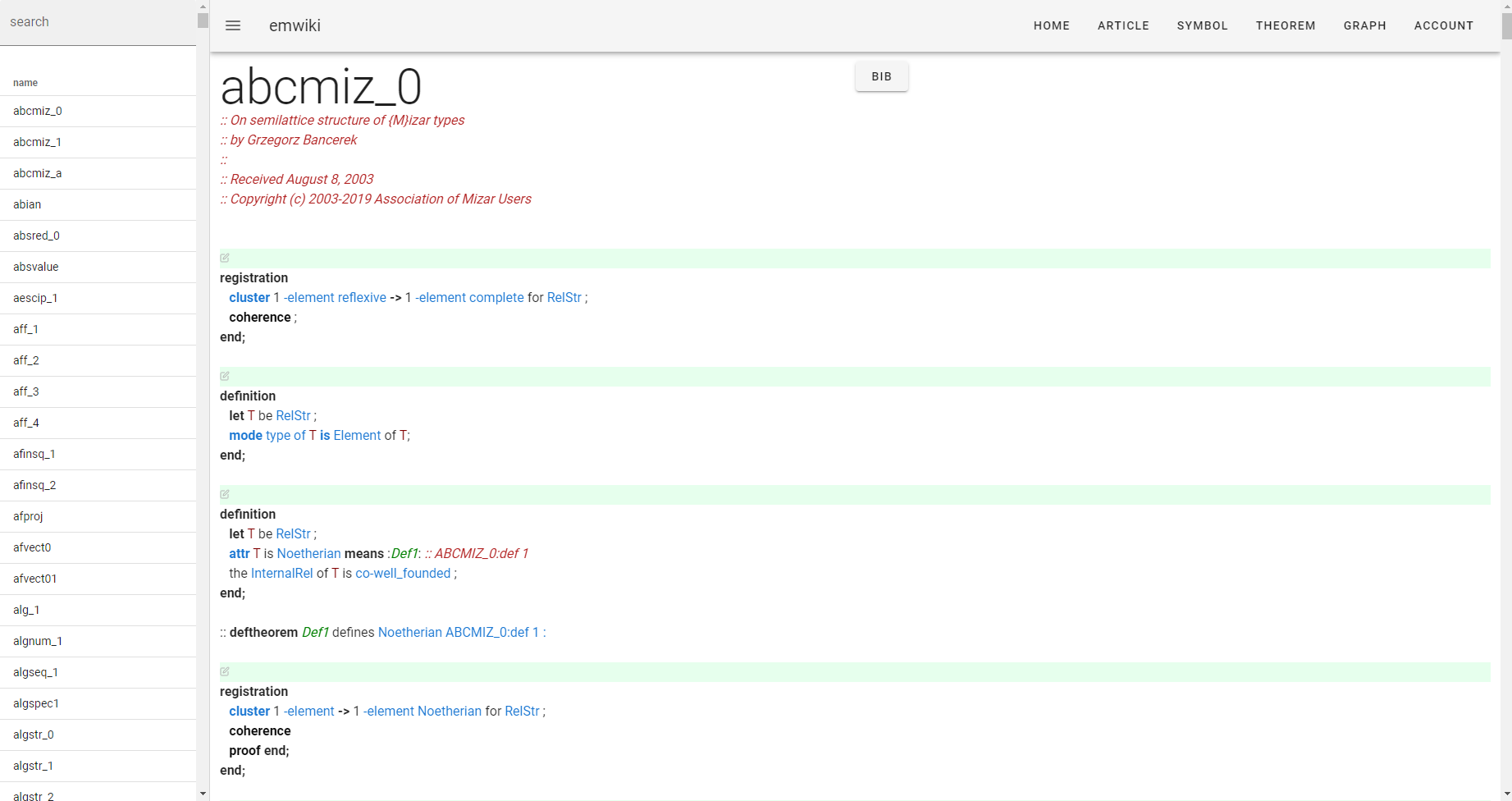}
    \caption{emwiki} \label{fig:emwiki}
\end{figure}

The screenshot of the Wiki feature is shown in Fig. \ref{fig:emwiki}.

\section{Search Function}
The emwiki system has three types of search components.
The two incremental search components accept the name of articles and symbols as input.
As of 2018, the MML contains 1,290 articles and 8,852 symbols~\cite{bancerek2018role}, and being able to search for these articles and symbols efficiently is critical for users.
MML Reference~\cite{nakasho2015documentation} generates HTML documents from the MML to help users understand its symbols.
Each HTML page contains symbol definitions, referrers, and references.
It also has an incremental search function that enables users to search with a fast and intuitive operation.
We integrated MML Reference into emwiki and linked this search function with the Wiki function.

We also developed a flexible theorem search component.
For a long time, grep has been used for full-text search in the MML~\cite{matuszewski2005mizar}.
MML Query~\cite{bancerek2003information}, developed by G. Bancerek et al. in 2001, has dramatically improved the efficiency of the search of the MML.
However, MML Query is not easy for beginners, because it has its own query language.
MML Query is a pattern-matching-based search system.
Therefore, it is not good at searching modified (but logically equivalent) theorems.
The Alcor system~\cite{cairns2007integrating}, developed by P. Carins et al., provides an Latent Semantic Indexing (LSI)-based search function for the Mizar Mathematical Library.
We have implemented a similar LSI-based search component into the emwiki system.
The search component does not require its own query language, and the user can perform searches by inputting the desired theorems in the Mizar language.
We have also implemented a function to collect user evaluations by displaying a ``good'' button on the search results.
In the future, we plan to use the evaluations as training data for machine learning.

\section{Visualization of Library Dependencies}
The dependency graph component is designed to encourage library maintainers to refactor their libraries by visualizing the dependencies of files in the MML.
The MML has been maintained for the past 30 years by the University of Bialystok~\cite{grabowski2015four} and is continuously refactored using specialized tools to minimize the dependencies of the articles and eliminate the cyclical dependencies between groups~\cite{naumowicz2015tools}.
However, some of the Mizar language specifications hinder the refactoring of the MML~\cite{rudnicki2003integrity}.
The Mizar language does not have namespace or package features.
Also, the length of the Mizar file name is limited to 8 characters due to the legacy of MS-DOS constraints.
Therefore, all of the 1,290 articles with ambiguous file names are currently arranged in a flat structure in the MML.
Furthermore, the Mizar language has a constructor overload feature and a context-sensitive grammar in which the last imported constructor takes precedence.
These language specifications should be improved in the future, but we must continue to maintain our vast library of assets in the meantime.

J. Alama (2011)~\cite{alama2011mizar} created a database of reference relations among theorems, definitions, and notations in the MML and provided a way to access the dependency graph through a Web interface.
J. Heras et al. (2014)~\cite{heras2014hott} provides a tool to visualize the dependency graph of the HoTT library written in Coq.
J. Alama et al. (2012)~\cite{alama2012dependencies} extracted the dependencies in the Coq and Mizar libraries and performed a quantitative comparative analysis of the library features.
R. Marcus et al. (2020)~\cite{marcus2020tgview3d} proposed the TGView3D system, which renders the dependencies of formalized mathematical libraries as 3D graphs in a hybrid of force-directed and hierarchical layouts.

In this study, we adopted the hierarchical graph.
The hierarchical graph is intended to clarify the flow of dependencies and utilize refactoring, such as minimizing dependencies and eliminating circular references among groups.
In the Mizar language, reference relationships between articles are described in the environment part at the beginning of an article.
Since Mizar does not recursively load external files written in environment sections, a transitive reduction is performed to cut redundant edges before constructing the dependency graph.
The data structure is saved in dot and sfdp formats of Graphviz and is drawn in a Web browser using Cytoscape.js library.
The component also provides functions such as highlighting nearby nodes, searching nodes, moving nodes, hyperlinking to articles, and zooming in/out.
A detailed usage of the dependency graph component can be found in the Help menu in the upper right corner.

\section{Conclusions and Future Work}
In this paper, we have developed the emwiki system as a Web platform for hosting the MML.
This system is differentiated from existing services hosting mathematical libraries in consistency, extendability, and interoperability.
One of the most severe difficulties in the long-term use of a service that hosts formalized mathematical libraries is keeping up with library updates.
In this study, we attempted to solve this problem using Git and its merge function to maintain reference relationships during version updates.
Since this platform is built using Django, a new component development is stylized as adding Django applications.
Also, all components are allowed to access their version-controlled library, user management function, and database.
The components implemented on the platform share Web pages and data with each other.
For example, MML Reference, now integrated into the system, shares a Wiki function and Web pages.
Ease of operation is also improved through automated testing during development and deployment using containers.

For the dependency graph component, several issues remain.
While our hierarchical graph could properly place the dependencies of the articles, it did not classify the articles into meaningful groups.
Mizar articles have metadata such as the Mathematics Subject Classification (MSC)\footnote{{https://msc2020.org/}}.
We consider taking this information into account for graph layout and including it as additional information for article search.
We also think it will be effective to analyze the vocabulary contained in the articles and reflect the semantic classification in the layout algorithm.
Another beneficial enhancement is drawing theorem graphs.
However, handling as many nodes as the theorems in the MML requires a more powerful graph database and drawing library than the conservative library we currently use.

This project aims to improve the browsability, searchability, and comprehensive understandability of formalized mathematical libraries.
Although the system currently supports only the Mizar language, it could be extended to other languages.
It is also significant to incorporate other components, including a Web IDE.

It is also important to establish a cooperative framework for the long-term operation of the system. In particular, we need to seek the cooperation of the core team that maintains the MML and the Mizar system to create a flow to ensure consistency between the MML versions. Also, we would like to invite a wide range of developers from the Mizar community to work with us in defining requirements and developing new functions.

\section*{Acknowledgments}
This work was supported by JSPS KAKENHI Grant Number JP20K19863.

\bibliographystyle{splncs04}
\bibliography{main}
\end{document}